\documentclass{elsart}
\usepackage{graphicx}
\usepackage{amssymb}
\usepackage{amsmath}

\journal{arXiv}

\begin{document}

\begin{frontmatter}

\title{A volume-based description of gas flows with localised mass-density variations}
\author{S.\ Kokou Dadzie},
\ead{kokou.dadzie@strath.ac.uk}
\author{Jason M.\ Reese\corauthref{cor}}
\ead{jason.reese@strath.ac.uk} \and
\corauth[cor]{Corresponding author.}
\author{Colin R.\ McInnes}
\ead{colin.mcinnes@strath.ac.uk}
\address{Department of Mechanical Engineering, University of Strathclyde,
\newline Glasgow G1 1XJ, UK}

\begin{abstract}
We reconsider some fundamental aspects of the fluid mechanics model,
and the derivation of continuum flow equations from gas kinetic
theory. Two topologies for fluid representation are presented, and a
set of macroscopic equations are derived through a modified version
of the classical Boltzmann kinetic equation for monatomic gases. The
free volumes around the gaseous molecules are introduced into the
set of kinetic microscopic parameters. Our new description comprises
four, rather than three, conservation equations; the classical
continuity equation, which conflates actual mass-density and
number-density in a single equation, has been split into a
conservation equation of mass (which involves only the classical
number-density of the gaseous particles) and an evolution equation
purely of the mass-density (mass divided by the actual volume of the
fluid). We propose this model as a better description of gas flows
displaying non-local-thermodynamic-equilibrium (rarefied flows),
flows with relatively large variations of macroscopic properties,
and/or highly compressible fluids/flows.
\end{abstract}

\begin{keyword}
gas kinetic theory \sep Boltzmann equation \sep compressible fluids and flows
\sep Navier-Stokes equations \sep rarefied gas dynamics
\end{keyword}
\end{frontmatter}

\section{Introduction}
In rarefied gas dynamics, the Boltzmann kinetic equation is accepted
as describing the evolution of the gaseous particle distribution
function. This equation is presumed to be valid for any dilute gas
flow, and various kinetic models have been developed as approximations
to it. The most well-known are the BGK relaxation model
\cite{PhysRev.94.511}, the Chapman-Enskog perturbation method
\cite{chapman}, the Linearized Boltzmann equation with polynomial
decompositions \cite{cercignaniblanc}, and the Grad moments method
with Maxwellian weighting function \cite{grad-1949}, as well as more
recent extensions and developments of these models. All these
kinetic models lead to the set of three hydrodynamic equations,
usually called the Navier-Stokes equations, that have amply
demonstrated their success in describing typical engineering flows
with relatively small density variations. However, the description
of flows beyond the broad range of applicability of the
Navier-Stokes model (such as hypersonic flows, or micro- or
nano-scale gas flows) remains an active area of investigation.

The appropriateness of the Navier-Stokes model for a gas flow rests
on there being sufficient local homogeneity in the flow such that
macroscopic variables (e.g.\ mass-density, temperature, pressure
etc.) are locally relatively uniform, with only small departures
from thermodynamic equilibrium. In this case, a local Maxwellian
representation can be assumed in approximating the solution to the
Boltzmann equation. Departures of the real distribution function of
the gaseous particles from the local Maxwellian are assumed to be
only slight. This assumption is acceptable if the following condition is
satisfied by the macroscopic flow properties:
\begin{equation}
\label{localcond}
    \left| \frac{d \Phi}{\Phi} \right| \ll 1 ,
\end{equation}
where $\Phi$ is a flow macroscopic property. If condition
(\ref{localcond}) does not hold then one or more macroscopic
properties admit large relative variation, and the distribution
function cannot be safely assumed to depart only slightly from a
local equilibrium. The above assumptions and condition
(\ref{localcond}) play a key role in the various kinetic models.
Flows with large relative variations in their macroscopic properties
are not well understood (see, e.g.,
\cite{LockerbyReeseGallis2005b,reesegallislockerby2003}) although
there have been numerous attempts to develop hydrodynamic models for
this class of flows: see, e.g.,
\cite{koga1954-Chem.Phy,struchtrupbook,myong2004,jinslemrod2001}.

In this paper we introduce a new treatment for fluids and flows that
are sensitive to relative variations of macroscopic flow properties.
We first examine the topological representations of gases and
thereby attempt a rigorous definition of both ``mass-velocity'' and
``volume-velocity''. We then derive from the fundamental kinetic
theory the set of conservation equations corresponding to a
volume-representation of fluids. While this approach owes much to
the recent work of Howard Brenner
\cite{Brenner-Phys.vol2005,Brenner.PhysicaA.revs.2005} and Hans
Christian \"Ottinger \cite{ottingerbook,bardowottinger2006}
questioning the conventional fluid mechanical description, our focus
here is on putting the volume-representation of gases on a more
rigorous footing that is rooted in the fundamental kinetic theory.

\section{The topology of fluid representation}
\label{sec01} A fluid is composed of a great number of molecules
occupying a given volume in physical space. In the case of a gas,
the physical volume occupied by the gaseous molecules can be
regarded as the envelope of the physical domain occupied by the gas,
and this can be represented geometrically. In a fixed reference
frame this envelope may vary: for example, the gas may expand or
contract.

In this section we present two different topological spaces for
representing gases. The first is based on the individual point-mass
particles, and the second representation is based on the geometrical
envelope of the gas in physical space (i.e.\ the volume in which a
number of gaseous molecules is dispersed). These two topological
spaces are two basically different measures (or integration forms).
Associated with each are essentially different ways to perform
balances.

\subsection{``Mass-based representation'' and the measure defined by
 \newline an element of mass}
\label{secmesmass} A gas may be regarded as discrete point-mass
particles (the gaseous molecules) distributed in physical space.
Disregarding the shape of the geometrical domain occupied by the
group of particles, we are only concerned with the number of
point-mass particles.

A set, $\Xi_m$, is defined by the discrete (numerated) molecules.
Any open subset, $\Omega$, represents a prescribed number of
molecules. The set of whole subsets, $\Omega$, of $\Xi_m$ will be
denoted $\Lambda_m$. We may define the following application:
\begin{eqnarray}
\label{mes_mas1}
\Lambda_m &\rightarrow &  \mathcal{R^+}  , \\
\nonumber \Omega  &\mapsto & \mathrm{Mes}_m ( \Omega ) ,
\end{eqnarray}
where $\mathrm{Mes}_m (\Omega)= M \times n_{\Omega}$, with $M$ the
molecular mass and $n_{\Omega}$ the number of point-mass  molecules
represented by $\Omega$. Then $\mathrm{Mes}_m (\Omega )$ is the
total measurable mass of the molecules contained in $\Omega$. This
application defines a measure on the topological space
$(\Xi_m,\Lambda_m)$. For any two different open subsets, $\Omega_i$
and $\Omega_j $ in $\Lambda_m$, so that $\Omega_i \cap \Omega_j = \O
$, we have
\begin{equation}
\label{mes_mass2}
\mathrm{Mes}_m(\Omega_i) + \mathrm{Mes}_m(\Omega_j) = m_i + m_j \ ,
\end{equation}
where $m_i$ (or $m_j$)  is the mass constituting the element $\Omega_i$ (or $\Omega_j$).

\subsection{``Volume-based representation'' and the measure defined by
\newline an element of volume}
\label{secmesvol} 
\begin{figure}
\begin{center}
\includegraphics{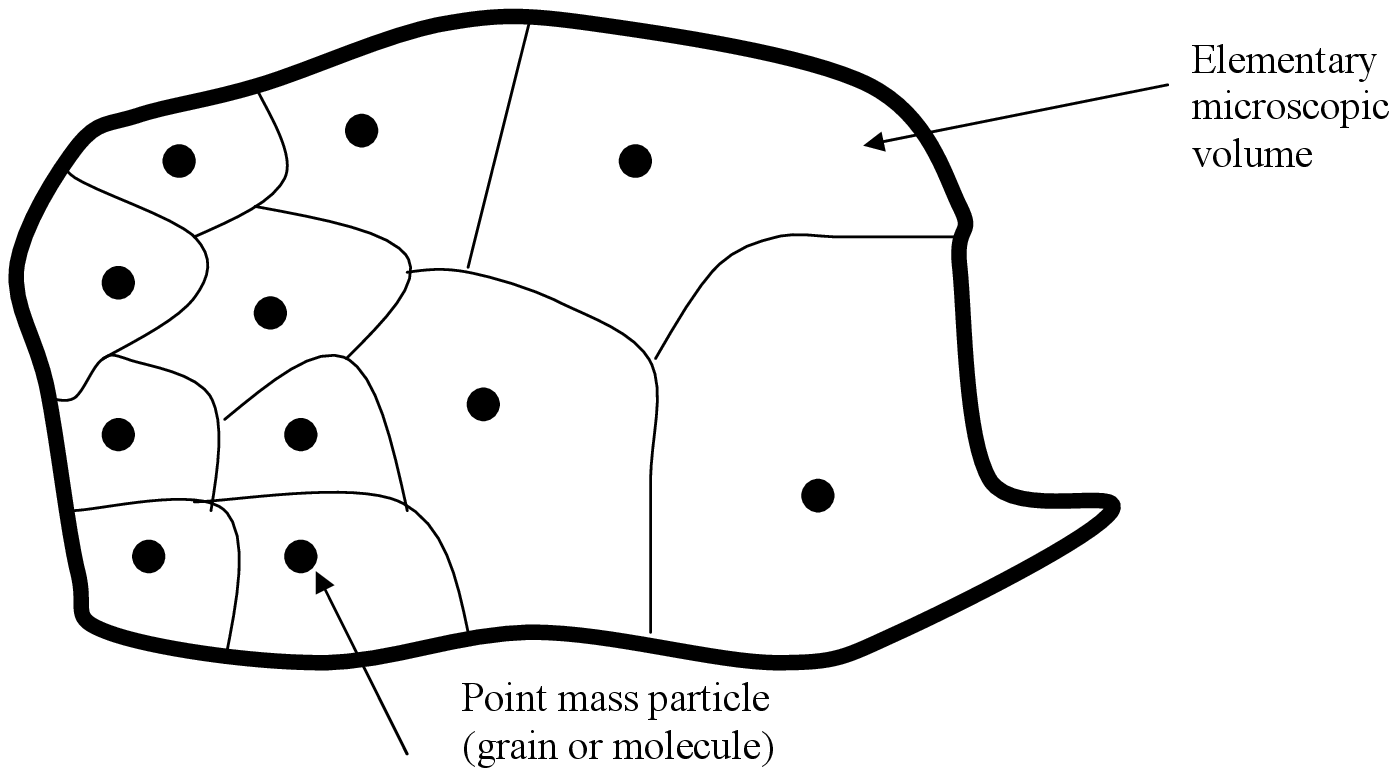}
\caption{Schematic (essentially, a Voronoi diagram) of microscopic
fluid volume distribution around a number of gaseous particles. The
elementary microscopic volume is small in regions with a greater
number of molecules, and large in regions with fewer molecules. A
``volume of fluid'' is represented by the physical domain in which
the molecules are dispersed.} \vspace{1.5em} \label{gas-schema}
\end{center}
\end{figure}

Instead of regarding only the point-mass
particles, we might be concerned with the physical domain occupied
by these dispersed particles. A prescribed number of molecules occupy
a given volume in physical
space. This volume can be regarded as the geometrical envelope of
this group of molecules. Considering a gas in motion at a given time,
$t$, the domain occupied by the group of gaseous molecules can be
split among the point-mass particles according to their real
position at this time $t$ (see figure \ref{gas-schema}). So each
gaseous particle is attributed a microscopic
fractional volume of the total volume occupied by the group of
molecules.

As a ``volume of fluid'' means the physical volume occupied by an
ensemble of gaseous molecules, the microscopic volume attributed to
each particle in this description can be read as a ``microscopic
volume of fluid''. Each microscopic volume contains only one
molecule, and the whole volume of fluid is given by the summation
over all microscopic volumes. In this second fluid topological
description, a prescribed ``amount of fluid'' or a ``volume of
fluid'' means ``a number of microscopic volumes''.

A set, $\Xi_v$, is defined by the microscopic volumes. Any open
subset, $\Omega$, represents a prescribed number of microscopic
volumes. The set of these subsets $\Omega$ of $\Xi_v$ will be
denoted $\Lambda_v$. We may define the following application:
\begin{eqnarray}
\label{mes_vol}
\Lambda_v & \rightarrow  & \mathcal{R^+}   , \\
\nonumber \Omega & \mapsto & \mathrm{Mes}_v ( \Omega ) ,
\end{eqnarray}
where $\mathrm{Mes}_v ( \Omega)$ is the measurable value of the
geometrical volume represented by $\Omega $. This application
defines a measure on the topological space $(\Xi_v,\Lambda_v)$. It
should be noted that, according to the microscopic volume
distribution (see also figure \ref{gas-schema}), the elementary
microscopic geometrical volumes are such that, for a couple of
microscopic volumes $\Omega_i $ and $ \Omega_j$,
\begin{equation}
\label{mes_vol1}
\Omega_i \cap \Omega_j = \O  .
\end{equation}
This ensures the measure additivity property. Therefore, for two
different open subsets, $\Omega_i$ and $\Omega_j $ in $\Lambda_v$, so that
$\Omega_i \cap \Omega_j = \O $, we have
\begin{equation}
\label{mes_vol2}
\mathrm{Mes}_v(\Omega_i) + \mathrm{Mes}_v(\Omega_j) = v_i + v_j  ,
\end{equation}
where $v_i$ (or $v_j$) is the measurable value of the volume element
$\Omega_i$ (or $\Omega_j$).

We note that:
\begin{itemize}
\item from a purely mathematical point of view, an ``elementary amount of
volume'' in
the second description plays the same role as an ``elementary amount of
mass'' in the first description;
\item figure \ref{gas-schema} may also be regarded as representing the
element termed a
``fluid particle'' in fluid mechanics, which is a volume domain containing a
great number of molecules.
\end{itemize}

\subsection{The two barycentric velocities: volume velocity and mass velocity}
\begin{figure}
\begin{center}
\includegraphics{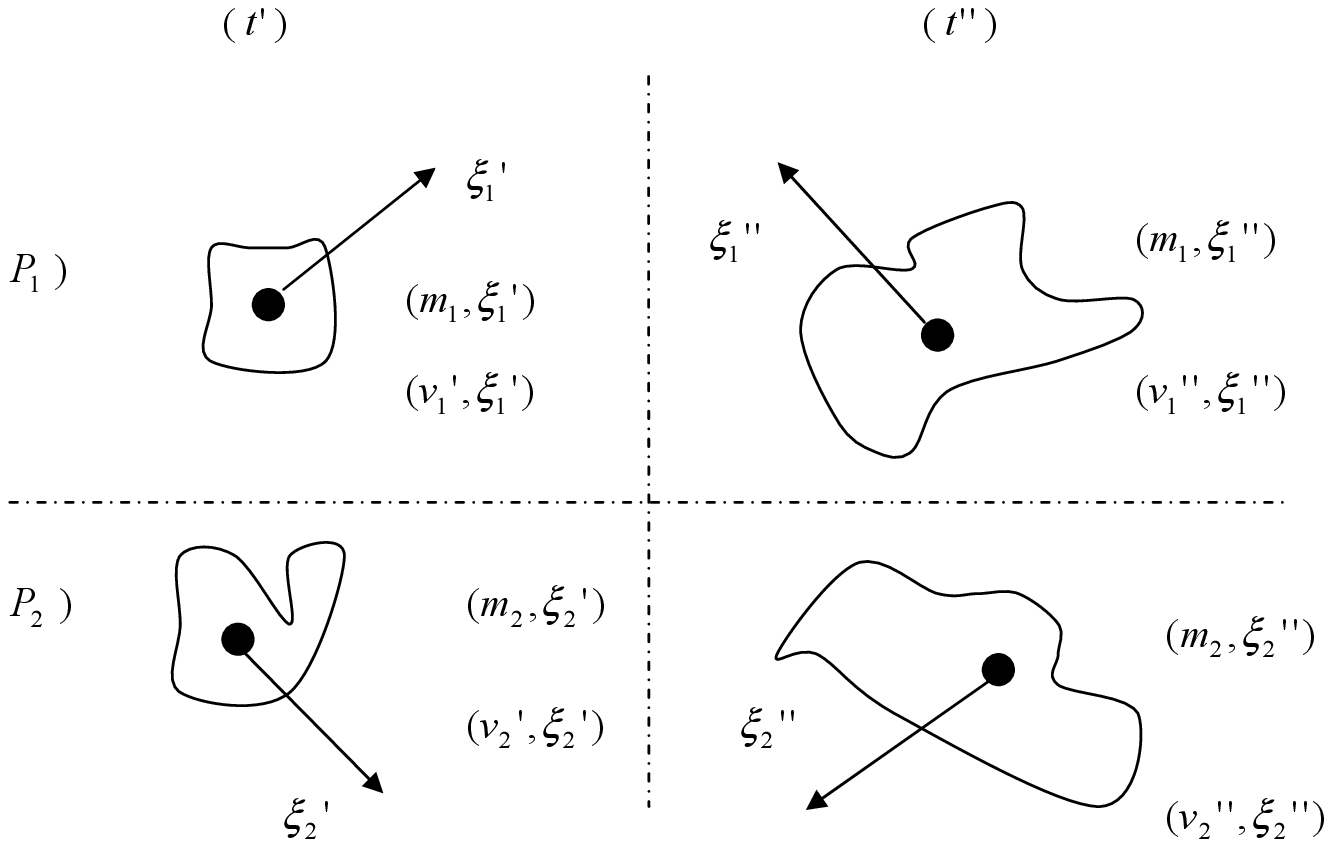}
\caption{Schematic of ``mass barycentric points'' and ``volume
barycentric points''.
The elements represented in this figure are to be taken as individual
elements from
figure~\ref{gas-schema}.}
\vspace{1.5em}
\label{shema-vitesse}
\end{center}
\end{figure}

Consider the motion of two molecules $P_1$ and $P_2$ between time
$t^{\prime}$ and $t^{\prime\prime}$. At time $t^{\prime}$, the mass
of molecule $P_1$ is $m_1$ and the mass of $P_2$ is $m_2$. Similarly
the microscopic volume associated with $P_1$ is $v_1$ while that
associated with $P_2$ is $v_2$. Each molecule has a given velocity
at each time and at each position (see figure \ref{shema-vitesse}),
which will be denoted  $\xi$.

From a mathematical point of view, two barycentric point systems can
be perfectly defined according to the mass of the particles and
according to the microscopic volumes associated with the particles.
In so doing, we have on one side a system of barycentric points
represented by the couples $(m_i,\xi_i)$ and on the other side a
system represented by the couples $(v_i,\xi_i)$. These two
barycentric systems lead to two different barycentric velocities (or
mean velocities).

More precisely, considering the molecules $P_1$ and $P_2$, the
velocity associated with both molecules, when regarding the mass, is
at time $t^{\prime}$ related to $m_1 \xi_1^{\prime} +
m_2\xi_2^{\prime}$ and at time $t^{\prime\prime}$ related to $m_1
\xi_1^{\prime\prime} + m_2\xi_2^{\prime\prime}$: the mass of each
molecule does not change in time. On the other hand, when regarding
the volume, the velocity associated with both molecules' motion is
at time $t^{\prime}$ related to $v_1^{\prime} \xi_1^{\prime} +
v_2^{\prime}\xi_2^{\prime}$ and at time $t^{\prime\prime}$ related
to $v_1^{\prime\prime} \xi_1^{\prime\prime} +
v_2^{\prime\prime}\xi_2^{\prime\prime}$: the microscopic volume does
change in time.

These two barycentric velocities are evidently different, as volume
and mass are two different weighting elements. The volume barycentric
velocity accounts for the variations of the microscopic volumes
in time and space, in addition to the variations of the velocities
of the point-mass particles. However, the mass barycentric
velocity accounts only for the variations of the velocities
of the point-mass particles, since the mass of the molecules is constant.

The two barycentric velocities will only be equivalent if there
is a linear relation between mass and microscopic volumes,
which would mean a uniform distribution of the particles in the fluid
without variation of the global volume occupied the particles.

The barycentric velocities are expressed:
\begin{itemize}
\item  mass barycentric velocity, $U_m$,
\begin{equation}\label{defmassvelocity}
\mathrm{Mes}_m (\Omega_m) \cdot U_m = \int_{\Omega_m} \xi d_M ,
\end{equation}
\item volume barycentric velocity, $U_v$,
\begin{equation}\label{defvolvelocity}
\mathrm{Mes}_v (\Omega_v) \cdot U_v = \int_{\Omega_v} \xi d_V  ,
\end{equation}
\end{itemize}
where the integral summations are founded by the measures defined in
previous sections, and related to the mass-based and the
volume-based representations, respectively.

We note here that the definitions (\ref{defmassvelocity}) and
(\ref{defvolvelocity}) correspond to what have been termed recently
``mass-velocity'' and ``volume-velocity'', respectively
\cite{Brenner-Phys.vol2005,Brenner.PhysicaA.revs.2005}. The
distinction between these mean velocities can also be shown from
generalized concepts
of ``centre of mass'' in more sophisticated differential geometries
in physics \cite{pavsic-arxiv2003}.

\subsection{Hydrodynamic velocities and the equations of fluid mechanics}
\label{secvites-hydro} In contrast to the motion of a solid body,
hydrodynamics involves the motion of a ``volume of fluid''. The
motion of a volume occupied by gaseous molecules is founded on the
concept of an element termed a ``fluid particle''. Accordingly, the
mean velocity of any volume of fluid should be
defined by the mean velocity obtained by weighting each microscopic
velocity by the respective microscopic volume of fluid.

In other words, and referring again to figure \ref{gas-schema}, the
mean velocity of a fluid particle, which is an element of volume, is
not the mean mass-velocity given by the mass-based representation,
but the velocity obtained by weighting the microscopic velocities
with the microscopic volumes. This is because the mean velocity
obtained by using mass as the weighting element corresponds strictly
to the velocity of a total amount of mass (the centre of mass),
independent of the volume of space occupied and therefore
independent of the volume of the fluid particle.

While the velocity of the fluid
particle should properly be attributed to the volume-based barycentric
velocity, the choice of hydrodynamic velocity
is not, however, the sole questionable point in the process of
deriving the fluid equations. The main point lies in being
consistent in the choice of velocity when performing balances for
mass, momentum and energy. This will be addressed below.

\subsection{An uncertainty in classical fluid mechanics topology}
\label{seccriticism} The two measures described in sections
\ref{secmesmass} and \ref{secmesvol} are connected to two different
ways of performing integrations and balances. In the derivation of
the conventional set of fluid mechanics equations, known as the
Navier-Stokes equations, a clear distinction has not been made
between mass-based integrations and volume-based integrations, and
the volume barycentric velocity and the mass barycentric velocity
are treated as equivalent.

If $V$ is a control volume of fluid, and $m_V$ the total mass
contained in $V$, the total amount of any property $Q$ (momentum or
energy) carried by the control volume (which is, strictly speaking,
the total amount of $Q$ carried by the amount of matter $m_V$) is
generally given by
\begin{equation}
\label{correctinteg}
\int_{\Omega_{m_V}} Q d_M \ ,
\end{equation}
which simply means the summation over all the elementary quantities
of $Q$ carried by the elementary point-mass particles contained in
the volume $V$.

In classical fluid mechanics, for instance when deriving
hydrodynamic equations such as the Navier-Stokes equations, the
amount of property $Q$ carried in $V$ is usually taken as
\begin{equation}
\label{classicinteg}
\int_V  \rho Q d_V \ ,
\end{equation}
where $\rho$ is taken as the fluid mass-density. From a mathematical
point of view, the total amount given by expression
(\ref{correctinteg}) corresponds to the total amount given by
expression (\ref{classicinteg}) if $\rho$ is uniform over the
control volume, $V$, i.e.\ that $d_M =\rho d_V$. Both expressions
are equivalent only if there exists a simple linear relation between
the quantity of mass and the quantity of volume.

It is common in fluid mechanics to argue expression
(\ref{classicinteg}) through the assumptions that a fluid particle
contains a great number of molecules, and that these molecules are
uniformly distributed so that $\rho$ can be viewed as being locally
constant. However, these classical assumptions cannot be said to be
completely fulfilled in some important situations, so a perfect equality
between expressions (\ref{correctinteg}) and (\ref{classicinteg})
cannot always be presumed. Balances based on expression
(\ref{classicinteg}) seem to be doubtful when (strong) density
gradients are present in the fluid, or in sufficiently rarefied
gases where the fluid volume must be quite large to incorporate
enough molecules.

To ensure equivalence of expressions (\ref{classicinteg}) and
(\ref{correctinteg}) from a statistical point of view, we suggest
replacing $\rho$ in expression (\ref{classicinteg}) with the mean
value, $\left\langle\rho\right\rangle$, of $\rho$ over the fluid
particle:
\begin{equation}
\label{classicinteg2}
\int_V  \left\langle \rho \right\rangle Q d_V \ .
\end{equation}
But in this situation, two points need to be emphasised. First, the
question of the definition of the mass-density arises, because
$\rho$ cannot be simply defined as a limiting value of the
mass-to-volume ratio \cite{chapman}. The mass-density definition
introduced in expression (\ref{classicinteg}) is properly defined only
when there are sufficient molecules in the fluid particle to smooth
out any density fluctuations. Second, if $Q$ is set to be the
molecule velocities then expression (\ref{classicinteg}) or
(\ref{classicinteg2}) embody the equality of ``volume barycentric
velocity'' and  ``mass barycentric velocity''. That is to say, for a
fluid element,
\begin{equation}
\int_{\Omega_m} \xi d_m = \int_V \left\langle \rho \right\rangle \xi d_V
=  \left\langle \rho \right\rangle
 \int_{\Omega_v}  \xi d_V  ,
\end{equation}
then a fluid particle velocity is such that
\begin{equation}
U_m =  U_v  .
\end{equation}
Obviously, this last equality is mathematically valid only for
uniformly distributed molecules, or in the event of equivalence
between the two measures described in sections \ref{secmesmass} and
\ref{secmesvol}.

The classical fluid mechanics argument that supports expression
(\ref{classicinteg}) is also the local thermodynamic equilibrium assumption in
gas dynamics. Assumption of local thermodynamic equilibrium is acceptable when
there are no large variations in the thermodynamic
parameters in the whole fluid domain concerned. In this case,
expression (\ref{classicinteg}) would be valid, as density may be safely
considered as locally constant. The gas flow Knudsen number is usually defined
as
\begin{equation}
\label{knudsennumber}
K_n= \frac{\lambda_m}{\rho}\frac{\delta \rho}{\delta x}  ,
\end{equation}
where $\lambda_m$ is the molecular mean free path and $x$ a spatial
coordinate. Local equilibrium is assumed when $K_n \rightarrow 0$,
which is the range admitted for validity of the classical
Navier-Stokes hydrodynamic equations. The equivalences that we
question in this section of the paper are only acceptable in the range $K_n
\rightarrow 0$. Therefore it remains to investigate how to
incorporate local non-uniformity into hydrodynamic models of fluid
flows.

\subsection{A paradox in the classical continuity equation}
\label{secparadox} Let us consider a mass, $M_0$, of some rarefied
gas containing a given number, $N_0$, of monatomic gaseous particles
occupying a volume, $V_0$. The particles are uniformly distributed
in $V_0$. Suppose that at time $t=0$ this gas is placed in a vacuum
(with no boundaries and no external force applied). The only forces
in the gas are through the interaction potential between particles.
We are interested in the time evolution of the mass-density of the
fluid.

As the molecules will scatter in all directions, it is expected that
the mean mass-velocity will be zero, and the classical continuity
equation predicts only a constant mass-density. However, the
mass-density of the fluid (i.e.\ mass divided by volume of the gas)
would be expected to decrease in time since the gas volume will
grow. Evolution of the mass-density without a mass-velocity is in
contradiction to the classical continuity equation.

Anticipating the discussion below and in Appendix
\ref{numb-dens-app} of the difference between number-density and
mass-density, another inconsistency in the classical continuity
equation is outlined in Appendix \ref{euler-contra}.

\section{Kinetic theory and mass-density variations}
In this section the problem of local density variations that we
have outlined in previous sections is addressed within the framework
of the kinetic theory of monatomic gases.

\subsection{The distribution function in classical kinetic theory }
Consider a physical space with reference to a fixed inertial frame
$(X_1,X_2,X_3)$. In classical kinetic theory, $f(t, X, \xi )$,
termed ``the particle distribution function'', is the probability
number-density of particles which, at a given time $t$, have their
velocities in the vicinity of the velocity $\xi$, and are located in
the vicinity of the fixed reference position $X$. This probability
density gives a number of particles at time $t$.

These particles occupy some geometrical domain, $v(t,X)$, in
physical space, which is the (local) volume occupied around the
fixed position $X$. But the distribution function as it stands does
not contain information about how these particles are scattered in
$v(t,X)$, or about the measurable value of $v(t,X)$.

We can see that the distribution function, $f(t, X, \xi )$, contains
the following information about the molecules at any time, $t$:
velocity $\xi$, momentum $M\xi$, energy $M\xi^2$.  None of the
parameters $t$, $X$, or $\xi$ contains information about the real
volume occupied by the group of particles. That is to say,
macroscopic descriptions derived from these parameters and from
$f(t, X, \xi )$ cannot contain information about the volume occupied
by the fluid.

The classical assumption about volume in kinetic theory is to
consider the space element $d_X= d_{X_1}d_{X_2}d_{X_3}$, which
defines the vicinity of the reference position $X$ and which is,
strictly speaking, a fixed element of space related to the frame
$(X_1,X_2,X_3)$. Then it is assumed that the volume occupied by the
particles is the fixed element $d_X$, and that the molecules are
uniformly distributed in this element \cite{chapman}.

Consequently, the classical distribution function, $f(t, X, \xi )$,
suffers from the same criticism presented in section
\ref{seccriticism}. Nor does the classical conception of a
distribution function admit evolution of the volume, $v(t,X)$,
occupied by the gaseous molecules in the fixed frame
$(X_1,X_2,X_3)$. The current conception of $f(t, X, \xi )$ is
suitable for a (locally) uniform dispersion of molecules, and for
gases with no large compressibility effects. It cannot correctly
treat gases flowing under higher compressibility (or rarefaction)
because in these cases the assumption that $d_X\equiv v(t,X)$ is
doubtful.

\subsection{Kinetic theory modified for local density variations}
According to our microscopic volume representation of figure
\ref{gas-schema}, a microscopic elementary volume is assigned to
each gaseous particle. Therefore, we may consider a distribution
function which incorporates the microscopic elementary volume as
follows:
\begin{quote}
\emph{$f(t, X,\xi, v )$ is the probability number-density of
particles which, at a given time $t$, are located in the vicinity of
position $X$, have their velocities in the vicinity of velocity
$\xi$, and have an assigned microscopic volume in the vicinity of
volume $v$.}
\end{quote}

It is important to note that what is termed here the ``vicinity of
reference position $X$'' is defined by a fixed element $d_X =
d_{X_1} d_{X_2}d_{X_3}$ that is related to the fixed reference frame
$(X_1, X_2,X_3)$ within which the fluid motion is being
investigated. This element must not be confused with the element of
volume $v$, which is defined through the real geometrical volume
envelope of the space occupied by an ensemble of molecules. In our
new $f(t, X, \xi, v )$, $v$ is the measurable positive value of the
geometrical microscopic volume, and quite distinct from $X$. The
microscopic volume $v$ can vary in an element of volume $d_X$: a
prescribed number of particles can reduce their volume space or
expand it whilst the element $d_X $ is kept fixed.

We note that, through our definition of $f(t, X,
\xi, v )$, the element $d_X$ may or may not contain a great number
of particles, and therefore may or may not contain a great number of
microscopic volumes of fluid, $v$.

As the origin of any motion is the individual motion of the
molecules (and not of the microscopic volumes!), the statistical
kinetic equation of the evolution of $f(t, X, \xi, v )$ can be
written similarly to the classical Boltzmann kinetic equation, i.e.,
\begin{equation}
\label{eq.boltzmann.kok}
\frac{\partial f}{\partial t} + (\xi \cdot \nabla ) f +(F \cdot
\nabla_\xi ) f + W  \frac{\partial f}{\partial v} = \int \int (f^*
f_1^* - f f_1) \sigma \xi_r d_{\omega} d_{\xi_1} ,
\end{equation}
where $f = f(t, X, \xi, v )$ and $f_1 = f(t, X, \xi_1, v_1 )$ refer
to post-collision particles, $f^* = f(t, X, \xi^*, v^* )$ and $f_1^*
= f(t, X, \xi_1^*, v_1^* )$ refer to pre-collision particles, $\xi_r
= \xi-\xi_1 $ is the particle relative velocity, $\sigma$ the
collision differential cross section, $d_\omega$ an element of solid
angle, and $\nabla_\xi$ denotes the formal operator $\nabla_\xi
=\partial/\partial \xi_x +\partial/\partial \xi_y +
\partial/\partial \xi_z $. We recall that the collision integral
(right-hand-side of equation \ref{eq.boltzmann.kok}), is based on
the elementary dynamics of collisions between two point-mass particles,
and uses their centre-of-mass.

The new term involving $W$ in equation (\ref{eq.boltzmann.kok})
arises from the introduction of the new variable $v$ into the
distribution function. In fact, $W$ may also be written $dv/dt$, and
then the supplementary term appears as the variation of $f$ due
purely to volume lost to, or gained from, external space. Like the
term in $F$, which represents the external body force contribution
to the variation of $f$, $W$ represents the contribution of any
volume change to the variation of $f$. Obviously, the rate of volume
variation, $W$, should not depend on the microscopic parameters, and
the variables $t$, $X$, $\xi$ and $v$ are independent. For example,
$W$ could be generated by macroscopic pressure gradients. In the
following, we will suppose no body force, i.e.\ $F=0$.

According to our volume representation, the microscopic volume of
two different particles involved in a collision is ($v + v_1$) after
the collision and ($v^* + v_1^*$) before the collision. This
microscopic volume will vary in time because the particle
repartition is different at each time. During a collision, the
microscopic volume carried by the particles is not affected by the
dynamics of the interaction. Consequently, the variation of volume
during collisions is only due to the variation of the microscopic
volumes of both particles in time. As we are considering
dilute gases, where the collision time is short compared to other
characteristic times, notably the time between two collisions, we
can assume that the variation of volume is larger during the
relatively long time between two collision than
during the short time of the collision itself. As a result,
the microscopic volumes can be considered as conserved
over the collision time, i.e.\
\begin{equation}
v^* + v_1^* = v + v_1 .
\end{equation}
Thus we have, during a collision, a set of four conserved
quantities: the microscopic volume, $v$, supplemented by the three
usual conserved quantities, i.e.\ mass, momentum and energy.

\subsubsection{Definitions of macroscopic quantities}
The local number-density of the molecules within the fixed reference frame
(i.e. referring to the element of volume, $d_X$) is given by:
\begin{equation}
n(t,X) =  \int_{-\infty}^{+\infty}  \int_0^{+\infty} f(t, X, \xi, v )  d_v d_\xi  \ ,
\end{equation}
while the local mean value, $\bar{Q}(t,X)$, of any property $Q$ in $d_X$ can be defined by:
\begin{equation}
\label{meanvalu}
n(t,X) \bar{Q}(t,X)   =  \int_{-\infty}^{+\infty}  \int_0^{+\infty} Q  f(t, X, \xi, v )
d_v d_\xi \ .
\end{equation}
For example, the local mean volume around each particle,
$\bar{v}(t,X)$, is defined by:
\begin{equation}
n(t,X) \bar{v} (t,X) = \int_{-\infty}^{+\infty}  \int_0^{+\infty}
v f(t, X, \xi, v )  d_v d_\xi  \ .
\end{equation}
From this mean value of the microscopic volume, a local mean value of the mass-density
in the vicinity of position $X$ can be properly defined through:
\begin{equation}
\label{massdensitydef}
\bar{\rho} (t,X)  = \frac{n(t,X) M }{n(t,X)\bar{v}(t,X)} =
\frac{M}{\bar{v}(t,X)} \ ,
\end{equation}
where $M$ is the molecular mass. The corresponding specific volume is given by
$\bar{v}/{M}$.

We note that $n\bar{v}d_X$ is the actual volume of fluid in the
vicinity of $X$, containing $nd_X$ gaseous molecules. In classical
kinetic theory, these $nd_X$ molecules are always assumed to occupy,
and be uniformly dispersed in, the fixed element of space $d_X$. Our
new description is evidently different: while the new definitions of
mass-density and specific volume represent mean values, and depend
on $t$ and $X$, the gaseous molecules need not necessarily be
uniformly dispersed in the vicinity of $X$. Moreover, the total
volume of fluid around position $X$ is $n\bar{v}d_X$, not $d_X$.

Appendix \ref{numb-dens-app} contains more details on distinguishing
between $n$ and $\bar{\rho}$.

Two mean velocities can be defined. First, the local mean mass-velocity,
$U_m (t,X)$, is given through
\begin{equation}
\label{vitessemass}
M n(t,X) U_m (t,X) = \int \int M  \xi f(t, X, \xi, v ) d_\xi d_v .
\end{equation}
A local volume-velocity, $U_v(t,X)$, can also be defined by,
\begin{equation}
\label{vitessevolume}
\bar{v} (t,X) n(t,X)  U_v (t,X) = \int \int  v \xi f(t, X, \xi, v ) d_\xi d_v .
\end{equation}
We note that the mean velocities defined through equations
(\ref{vitessemass}) and (\ref{vitessevolume}) are equivalent to the
two velocity definitions pointed out in earlier sections of this paper.

The classical particle peculiar velocity is defined through the
mass-velocity, i.e.\
\begin{equation}
\label{peculiarmass} C  = \xi - U_m \ .
\end{equation}
But another peculiar velocity can also be defined using the
volume-velocity $U_v$:
\begin{equation}
\label{peculiarvol} C^{\prime}  = \xi - U_v \ .
\end{equation}
The peculiar velocity given by equation (\ref{peculiarmass}), is
usually supposed to define the random motion of the point-mass
particles. However, it may be noted in equation (\ref{peculiarmass})
that only the macroscopic motion of the centre-of-mass, $U_m$, is
removed from the point-mass velocities inside the peculiar velocity
$C$. Therefore, macroscopic motions due to expansion or compression
of the fluid element may still be contained in $C$.

\subsubsection{Conservation equations from the modified kinetic equation}
We now show the derivation of new macroscopic equations from our
kinetic equation (\ref{eq.boltzmann.kok}). This procedure is similar
to the classical one: equation (\ref{eq.boltzmann.kok}) is
multiplied by the microscopic quantities $v$, $M$, $(M\xi)$,
$(M\xi^2)$ and then the result is integrated over $d_v$ and $d_\xi$.
In this process it should be kept in mind that $t$, $X$, $\xi$, and
$v$ are independent variables, while any mean value of a microscopic
quantity given through definition (\ref{meanvalu}) depends on $t$
and $X$.

\begin{itemize}
\item Conservation of volume. Multiplying equation (\ref{eq.boltzmann.kok}) by the microscopic
element $v$, and integrating over $v$ and $\xi$, we obtain,
\begin{equation}
\label{vtimeboltzmann}
\int \int v \frac{\partial f}{\partial t} d_v d_\xi + \int \int v
(\xi \cdot \nabla ) f d_v d_\xi   + W \int \int v \frac{\partial
f}{\partial v}d_v d_\xi = 0  ,
\end{equation}
where the collision integral term vanishes. Since $t,X,\xi $ and $ v$
are independent variables, this equation reduces to
\begin{equation}
\int \int  \frac{\partial v f}{\partial t} d_v d_\xi + \int \int
\nabla \cdot (v f \xi) d_v d_\xi + W \int \int v \frac{\partial
f}{\partial v}d_v d_\xi= 0 ,
\end{equation}
which can also be written,
\begin{equation}
\label{vtimeboltzmann2}
\frac{\partial n \bar{v}}{\partial t} + \int \int \nabla \cdot (v f
U_m) d_v d_\xi + \int \int  \nabla \cdot (v f C ) d_v d_\xi + W \int
\int v \frac{\partial f}{\partial v}d_v d_\xi= 0 .
\end{equation}
Using partial integration and the integrability condition,
$\lim_{v\rightarrow +\infty}(v f) = 0$, the third integral term in
relation (\ref{vtimeboltzmann2}) gives
\begin{equation}
W \int \int v \frac{\partial f}{\partial v}d_v d_\xi=- n W ,
\end{equation}
and relation (\ref{vtimeboltzmann2}) can then be written
\begin{equation}
\frac{\partial n \bar{v} }{\partial t} + \nabla \cdot [n \bar{v}
U_m] +
 \nabla \cdot [\int \int (v C f ) d_v d_\xi] -
nW = 0 .
\end{equation}
Finally, if we denote
\begin{equation}\label{Jv}
\mathbf{J}_v = \int \int (v C f ) d_v d_\xi ,
\end{equation}
then the first macroscopic equation obtained is an evolution
equation for the volume, and is written
\begin{equation}\label{volumediffuse}
\frac{\partial n \bar{v} }{\partial t} +  \nabla \cdot [n
\bar{v}U_m] +
 \nabla \cdot [\mathbf{J}_v] =nW .
\end{equation}
The quantity $\mathbf{J}_v$ in equation (\ref{Jv}) represents a flux
of volume due to point-mass particle random motions defined with the
peculiar velocity $C$.

From equations (\ref{vitessevolume}) and
(\ref{peculiarmass}), the local mean volume-velocity of the fluid
may be written:
\begin{equation}
n \bar{v}  U_v = \int \int  v U_m  f(t, X, \xi, v ) d_\xi d_v   +
\int \int v C  f(t, X, \xi, v ) d_\xi d_v .
\end{equation}
Consequently, the following relation between the volume-velocity and
the mass-velocity is deduced:
\begin{equation}
\label{relationvitesmassvol} n \bar{v}  U_v =  n \bar{v} U_m +
\mathbf{J}_v  .
\end{equation}

\item Conservation of mass. Multiplying equation (\ref{eq.boltzmann.kok})
 by the molecular mass $M$, and integrating over $v$ and $\xi$, we obtain:
\begin{equation}
\label{mtimeboltzmann}
\int \int M \frac{\partial f}{\partial t} d_v d_\xi + \int \int M
(\xi \cdot \nabla ) f d_v d_\xi + W \int \int M \frac{\partial
f}{\partial v}d_v d_\xi = 0  ,
\end{equation}
where the collision integral term vanishes. The third integral term
in equation (\ref{mtimeboltzmann}) is zero owing to the generalized
function character of $f$, i.e.\ $\lim_{v\rightarrow 0} f =0$ and
$\lim_{v\rightarrow +\infty} f=0$. The second macroscopic equation
obtained in this case is then:
\begin{equation}
\label{massconserv}
\frac{\partial M n}{\partial t} + \nabla \cdot [M n U_m] = 0 ,
\end{equation}
which is a typical equation of conservation of mass or, more
rigorously, conservation of the number of particles. Combining
equation (\ref{massconserv}) with the volume equation
(\ref{volumediffuse}) gives
\begin{equation}
\label{vollumediffuse}
n  \left( \frac{\partial \bar{v} }{\partial t} +  U_m \cdot \nabla
\bar{v}\right) +
 \nabla \cdot [\mathbf{J}_v] = nW  .
\end{equation}
Using the density $\bar{\rho}=M/\bar{v}$, this can be rewritten:
\begin{equation}
\label{vollumediffuse-masse}
\left( \frac{\partial \bar{\rho} }{\partial t} +  U_m \cdot
\nabla \bar{\rho}\right)
 -\frac{\bar{\rho}^2}{Mn }\nabla \cdot [\mathbf{J}_v] + \frac{\bar{\rho}^2}{M }W =0 .
\end{equation}

\item  Conservation of momentum. Multiplying equation (\ref{eq.boltzmann.kok}) by $M\xi$,
and integrating over $v$ and $\xi$, we obtain
\begin{equation}
\label{mxitimeboltzmann}
\int \int M\xi \frac{\partial f}{\partial t} d_v d_\xi + \int \int
M\xi (\xi \cdot \nabla ) f d_v d_\xi  + W \int \int M
\xi\frac{\partial f}{\partial v}d_v d_\xi= 0 ,
\end{equation}
where the collision integral term vanishes. As $t$ and $\xi$ are
independent variables, this equation can be written in the form:
\begin{equation}
\label{mxitimeboltzmann1}
\int \int \frac{\partial M f \xi }{\partial t} d_v d_\xi + \int \int
\nabla \cdot (Mf\xi_i \xi_j )
d_v d_\xi + W \int \int M \xi\frac{\partial f}{\partial v}d_v d_\xi= 0 ,
\end{equation}
where $\xi_i \xi_j $ is the second order tensor constituted by the
elements ($\xi_i \xi_j$). Here also, the third integral term in
relation (\ref{mxitimeboltzmann1}) is zero owing to the generalized
function character of $f$, i.e.\ $\lim_{v\rightarrow 0} \xi f =0$
and $\lim_{v\rightarrow +\infty} \xi f=0$. Then, using the
definition of peculiar velocity, we obtain the third conservation
equation:
\begin{equation}
\label{conservemomentum}
\frac{\partial M nU_m}{\partial t} + \nabla \cdot \left[
MnU_mU_m\right] +M \nabla \cdot \mathbf{P }  = 0 ,
\end{equation}
where $\mathbf{P}\equiv \mathbf{P}_{ij}(t,X)$ is the flux:
\begin{equation}
\mathbf{P}_{ij}(t,X)= \int \int (C_iC_j)f d_v d_\xi .
\end{equation}
We note that the conservation equation (\ref{conservemomentum}) is a
number-density (or mass) based equation and does not contain any
volume information, or the density $\bar{\rho}$. Using the mass
conservation equation (\ref{massconserv}), the momentum equation may
be written :
\begin{equation}
\label{conservemomentumnavier}
n \left(\frac{\partial U_m}{\partial t} + U_m \cdot \nabla U_m\right)
 + \nabla \cdot \mathbf{P}   = 0 .
\end{equation}

\item Conservation of energy. Multiplying equation (\ref{eq.boltzmann.kok}) by
$\frac{1}{2}M\xi^2$ and integrating over $v$ and $\xi$, we obtain
\begin{equation}
\label{mxixitimeboltzmann}
\int \int \frac{1}{2}M \xi^2 \frac{\partial f}{\partial t} d_v d_\xi
+ \int \int \frac{1}{2} M \xi^2 (\xi \cdot \nabla ) f d_v d_\xi + W
\int \int M \xi^2\frac{\partial f}{\partial v}d_v d_\xi= 0 ,
\end{equation}
which, following the independent variable properties, becomes
\begin{equation}
\int \int \frac{1}{2} M \frac{\partial f\xi^2}{\partial t} d_v d_\xi
+ \int \int  \frac{1}{2} M \nabla \cdot (\xi^2 f \xi) d_v d_\xi + W
\int \int M \xi^2\frac{\partial f}{\partial v}d_v d_\xi = 0 .
\end{equation}
Here also, owing to the properties of $f$, i.e.\ $\lim_{v\rightarrow
0} \xi^2 f =0$ and $\lim_{v\rightarrow +\infty} \xi^2 f=0$, the
third integral term vanishes. Using the peculiar and the mean
velocity definitions, we therefore have
\begin{eqnarray}
\label{energyconserve} \frac{\partial }{\partial t}
\left[\frac{1}{2} M n U_m^2 \right] && + \frac{\partial}{\partial t}
\left[ M n e_{in} \right]  +
 \nabla \cdot \left[\frac{1}{2}M n U_m^2U_m + M n
e_{in} U_m\right] \\ \nonumber && + \nabla \cdot  \left[M \mathbf{P}
\cdot U_m\right] + \nabla \cdot M \mathbf{q}= 0 ,
\end{eqnarray}
where we have introduced the quantity $e_{in}$ that is given
through:
\begin{equation}
\label{int-energy}
  n(t,X) e_{in}= \frac{1}{2} \int \int  C^2 f d_\xi d_v ,
\end{equation}
and the flux $\mathbf{q}(t,X)$, given by:
\begin{equation}
\label{enrgy-flux}
  \mathbf{q}(t,X) = \frac{1}{2} \int \int C^2 C f d_\xi d_v .
\end{equation}
Equation (\ref{energyconserve}) is the mean energy evolution
equation. Again, this equation is mass-based, and does not involve
any volume information. By using the mass conservation equation
(\ref{massconserv}), the energy equation may be rewritten :
\begin{eqnarray}
\label{energyconservenavier}
 n \frac{\partial }{\partial t} \left[\frac{1}{2} U_m^2 + e_{in} \right]
&& + n U_m \cdot \nabla  \left[\frac{1}{2} U_m^2 + e_{in}\right]
\\ \nonumber
&&
+ \nabla \cdot  \left[ \mathbf{P}\cdot U_m\right]
+ \nabla \cdot  \mathbf{q}= 0 .
\end{eqnarray}
\end{itemize}

\section{A new set of macroscopic conservation equations}
The set of macroscopic conservation equations obtained from our modified
kinetic equation (\ref{eq.boltzmann.kok}) are:
\begin{itemize}
\item equation (\ref{vollumediffuse}) for the volume;
\item equation (\ref{massconserv}) for the mass;
\item equation (\ref{conservemomentumnavier}) for the momentum;
\item equation (\ref{energyconservenavier}) for the energy.
\end{itemize}
In this set of equations, the mean velocity $U_m(t,X)$ is by
definition the mass-velocity in each equation; the various fluxes,
$\mathbf{J}_v(t,X)$, $\mathbf{q}(t,X)$, and $\mathbf{P}(t,X)$ are
the fluxes related to the peculiar velocity, $C$, defined in equation (\ref{peculiarmass}).

Our new set of macroscopic equations are rewritten below for
convenience, using the material derivative $D/Dt \equiv
\partial /\partial t + U_m \cdot~\nabla$
\begin{description}
\item[Continuity]
\begin{equation}
\label{massnewmacro} \frac{D n }{ Dt} = -  n \nabla \cdot U_m  ,
\end{equation}
\item[Mass-density]
\begin{equation}
\label{densitynewmacro} \frac{D \bar{\rho} }{ D t}
=\frac{\bar{\rho}^2}{M }\left [ \frac{1}{n }\nabla \cdot
[\mathbf{J}_v] - W \right] ,
\end{equation}
\item[Momentum]
\begin{equation}
\label{momentumnewmacro} n \frac{D U_m }{ D t}  = -  \nabla \cdot
\mathbf{P} ,
\end{equation}
\item[Energy]
\begin{equation}
\label{energymacro}
 n \frac{D }{ D t} \left[\frac{1}{2} U_m^2 + e_{in}\right]
= - \nabla \cdot  \left[ \mathbf{P}\cdot U_m\right] - \nabla \cdot
\mathbf{q} \ .
\end{equation}
\end{description}
The volume-velocity of the flow is related to the mass-velocity
through $ n \bar{v} U_v =  n \bar{v} U_m + \mathbf{J}_v$ .

\section{Discussion}
The new set of  macroscopic equations,
(\ref{massnewmacro})--(\ref{energymacro}), is a set of four
conservation equations instead of the usual three. The main novel
aspect of our set of equations is the replacement of the classical
continuity equation, which usually involves both the number-density
and the mass-density, by two separate equations: a pure conservation
equation of mass and an evolution equation for the mass-density.

Although the momentum and energy equations look similar to the
classical conservation equations, it should be noted that they are
mass-based equations through the number-density $n$, and they do not
involve directly the mass-density or the actual volume of fluid.
They also refer to a fixed inertial frame. This characteristic
matches well the interpretation of equation (\ref{momentumnewmacro})
as Newton's second law which is, in its correct form, a mass-based
law that disregards the actual volume that contains the mass.

In terms of new features introduced by the two new relations in the
set of equations, we refer to the difference between an
``incompressible flow'' and an ``incompressible fluid''. Recalling
that $\nabla \cdot U_m$ is the volume derivative following the
velocity $U_m$, the continuity equation (\ref{massnewmacro}) is
interpreted as the number-density variation due to the flow
compressibility. On the other hand, fluid compressibility effects,
that are due to the variation of the fluid mass-density, are
contained in relation (\ref{densitynewmacro}). The compressibility
involved in the new relation (\ref{densitynewmacro}) can be caused
by significant temperature or pressure variations in the fluid as it
flows, in contrast to the first form of compressibility effects
which are related to the flow speed or acceleration.

The classical fluid mechanics continuity equation combines
the mass-density and the number-density in a single equation,
which therefore assumes equivalency of compressibility effects arising
from the fluid and from the flow speed --- although it is well-known
experimentally that these two effects are different. Classifying a
flow as incompressible or not is difficult in the context
of the classical continuum equations
\cite{sharipo,morini-ICNMM2006,Gad-el-Hak1999.JFlu.Eng.}. However,
in our new model it is clear that a flow without
compressibility effects can only be assumed if there are
insignificant relative variations of the mass-density (through
equation \ref{densitynewmacro}) and insignificant volume variation
under acceleration, $\nabla \cdot U_m=0$ (through equation
\ref{massnewmacro}).

The departure of flow solutions of our new set of equations from
solutions of the classical equations may be expected in a range of
flows that suffer from high fluid compressibility effects, even if
the flow itself remains within the conventional incompressible
condition (i.e.\ the flow Mach number less than about $0.3$);
examples of such flows include those seen in and around microscale
devices \cite{Gad-el-Hak1999.JFlu.Eng.} and other rarefied gas flow
situations.

If there is no mass-density variation then there is no rate of
volume variation, i.e.\ $W=0$, and no flux of volume, i.e.\
$\mathbf{J}_v=0$. In this case, equation (\ref{densitynewmacro}) disappears
and our set of equations reverts to the usual three-equation
model in which the mass-density is simply $\bar{\rho} = M n$. On the
other hand, if the mass-velocity, $U_m$, is zero then equations
(\ref{volumediffuse}) and (\ref{relationvitesmassvol}) give,
\begin{equation}
\label{volumedynamic} \frac{\partial n \bar{v} }{\partial t} +
\nabla \cdot [n \bar{v} U_v] =nW \ ,
\end{equation}
which is thus an equation of expansion (or compression) of the
fluid in which $W$ is the term of volume of fluid production. This
equation accompanies the energy equation even in situations where
the mass-velocity evolution equation (\ref{momentumnewmacro}) can be
disregarded. An example of such a situation is the configuration
described as paradoxical in section \ref{secparadox}.

In our modified kinetic equation (\ref{eq.boltzmann.kok}), $W$
appears as the internal rate of change of volume occupied by the
gas, and is independent of the microscopic parameters, i.e.\ $W =d v
/dt$. If we assume this rate of change of volume is associated with
variations of the macroscopic parameters, such as the fluid
temperature and pressure which we denote in this paper as
$T^{\prime}$ and $p^{\prime}$, respectively, we can then write the
following formal expression,
\begin{equation}
\label{descr_W1} W = \frac{dv}{dt} =
\bar{v}\left(\frac{1}{\bar{v}}\frac{\partial v}{\partial
T^{\prime}}\right)_{p^{\prime}} \frac{d T^{\prime}}{dt} +
\bar{v}\left(\frac{1}{\bar{v}}\frac{\partial v}{\partial
p^{\prime}}\right)_{T^{\prime}} \frac{dp^{\prime}}{dt} \ +  ... \ ,
\end{equation}
or
\begin{equation}
\label{descr_W2} W  = \bar{v} \alpha \frac{dT^{\prime}}{dt} -
\bar{v}\chi \frac{d p^{\prime}}{dt} \ + ... \ ,
\end{equation}
with compressibility coefficients defined by:
\begin{equation}
\alpha = \left(\frac{1}{\bar{v}}\frac{\partial v}{\partial
T^{\prime}}\right)_{p^{\prime}}  , \hspace{2.5em} \chi = -
\left(\frac{1}{\bar{v}}\frac{\partial v}{\partial
p^{\prime}}\right)_{T^{\prime}} \ .
\end{equation}
The dimensions of these two coefficients suggest $\alpha \approx 1/
T^{\prime}$ and $\chi \approx 1/ p^{\prime}$. In this case, these
approximations represent simple equations to describe the variations
of the volume around gaseous particles as a function of the fluid
macroscopic parameters; they should not be taken as a derivation of
an equation of state. This is because equations (\ref{descr_W1}) and
(\ref{descr_W2}) concern variations of the ``microscopic volume''
around each particle, while a thermodynamic equation of state is
correctly applied only to a ``macroscopic volume'' of fluid --- and
requires equilibrium conditions.

Using the above approximation for $W$, the mass-density equation
(\ref{densitynewmacro}) can be rewritten:
\begin{equation}
\label{densitynewhydro3} \frac{D \bar{\rho}}{ D t} =
\frac{\bar{\rho}^2}{Mn }\nabla \cdot [\mathbf{J}_v] -
\bar{\rho}\left( \alpha \frac{d T^{\prime}}{d t} - \chi \frac{d
p^{\prime}}{dt}\right) \ .
\end{equation}
In this equation, the left-hand-side derivative is the material
derivative that involves $U_m$ while on the right-hand-side is a
total derivative which does not always equal the material
derivative. For example, the total derivative, $d \ /dt$, may be expressed using
velocity $U_v$ rather than the mass velocity $U_m$, i.e.
\begin{equation}
 \frac{d}{d t } =\frac{\partial }{ \partial t} +
 U_m\cdot \nabla   +  \frac{1}{n \bar{v}}\mathbf{J}_v  \cdot \nabla  \
 .
\end{equation}

Now, equation (\ref{densitynewhydro3}) enables us to solve in
outline the problem posed in section \ref{secparadox}. We admit zero
mass velocity in this flow configuration, so equation
(\ref{densitynewhydro3}) is re-written
\begin{equation}
\frac{1}{\bar{\rho} }\frac{\partial \bar{\rho}}{ \partial t} =
\frac{\bar{\rho}}{Mn }\nabla \cdot [\mathbf{J}_v] - \left( \alpha
\frac{d T^{\prime}}{d t} - \chi \frac{d p^{\prime}}{dt}\right) \ .
\end{equation}
As we are only concerned with the variation in time, we may assume
an approximately uniform evolution of the fluid domain, i.e $\nabla
\cdot [\mathbf{J}_v] = 0$ (i.e. the flux of volume depends only on
time, not space). Then the solution of this problem is determined
through
\begin{equation}
\label{dernier} \frac{1}{\bar{\rho} }\frac{\partial \bar{\rho}}{
\partial t} = - \left( \alpha \frac{d T^{\prime}}{dt} - \chi \frac{d
p^{\prime}}{dt}\right) \ .
\end{equation}
This solution is independent of the number-density, $n$ (which is
connected to the fixed reference frame of the observer), so the
solution of equation (\ref{dernier}) is independent of the reference
frame. As the compressibility coefficients do not depend explicitly
on time, but only on the gas properties such as temperature and
pressure, then the solution of equation (\ref{dernier}) can be
written:
\begin{equation}
\label{dernier2} \bar{\rho}  = \bar{\rho}_0 \exp\left\{ \alpha
(T^{\prime} - T^{\prime}_0) - \chi ( p^{\prime} -p^{\prime}_0)
\right\} \  ,
\end{equation}
with $\bar{\rho}_0\equiv\bar{\rho}_0(X)$, $p^{\prime}_0 \equiv
p^{\prime}_0(X)$ and $T^{\prime}_0 \equiv T^{\prime}_0(X)$ the
initial values of mass-density, pressure and temperature of the gas.
As a result, equation (\ref{dernier2}) gives the evolution of the
mass-density with the temperature and the pressure --- although this
does not affect the assumption of mass conservation embodied in
equation (\ref{massnewmacro}).

Equation (\ref{dernier}) is not completely new; it, or an
approximate form, is usually assumed in fluid mechanics for flows
presenting density variation effects. The originality of our model
is that this equation is actually embodied in the kinetic equation
(\ref{eq.boltzmann.kok}). This equation is therefore not a simple
phenomenological relation, as usually presented, but is contained
within the full set of conservation equations. Equation (\ref{dernier})
describes the
local mean value of the mass-density even when the classical
assumption of real local uniformity does not hold.

\section{Conclusions}
In this article we have suggested there may be inconsistencies in
the classical treatment of the ``mass of fluid'' and the ``volume of
fluid'' descriptions. These inconsistencies appear to become
important (a) if significant relative variations in density arise
in the fluid, and/or (b) in flows in which the local equilibrium
assumption does not hold. Two different representations of fluids
have been outlined, and a volume-based kinetic approach has
been introduced through a slightly modified version of the
Boltzmann kinetic equation.

The set of macroscopic equations derived from our modified kinetic
equation has a fundamental departure from the classical description
of fluid mechanics: an evolution equation purely of the mass-density
is added to the set of three conservation equations (for
number-density, momentum and energy). While conservation of mass is
embodied in an equation involving only the particle number-density,
mass-density evolution is embodied in a separate conservation
equation which invokes a flux of volume and incorporates parameters
that can generate fluid volume variation (such as temperature or
pressure variations). Our new model, therefore distinguishes between
compressibility effects arising from the ``fluid compressibility''
and from the ``flow compressibility''.

The nature of the velocity appearing in the classical Navier-Stokes
set of equations has recently been questioned
\cite{Brenner.PhysicaA.revs.2005}. It has been suggested, although
without rigorous proof, that in the set of fluid mechanics equations
the velocity to be used when writing the Newton viscosity law should
be the volume-velocity. In the present article we have shown that,
from the kinetic theory point of view, mass-velocity and
volume-velocity can be properly defined: mass-velocity gives only
the velocity of the centre-of-mass of a fluid element, while volume
velocity accounts for expansion or compression of the fluid
element.

In classical kinetic theory, the pressure tensor and heat flux are
systematically attributed to the fluxes $\mathbf{P}_{ij}(t,X)$ and
$\mathbf{q}(t,X)$ appearing in the conservation of momentum and
energy equations, and these fluxes are founded on the peculiar
velocity. But in the new volume-based kinetic approach introduced in
this paper, two different peculiar velocities can be defined.
Therefore further investigations are required in order to connect
the real pressure tensor, heat flux, and internal energy of the
fluid to the various fluxes appearing in the set of conservation
equations. We address this issue in a complementary paper, so that a
complete set of hydrodynamic equations is derived.

\section*{Acknowledgements}
The authors would like to thank Howard Brenner of MIT (USA), Gilbert
M\'eolans of the Universit\'e de Provence (France), and Chris
Greenshields of Strathclyde University (UK) for useful discussions.
This work is funded in the UK by the Engineering and Physical
Sciences Research Council under grant EP/D007488/1, and through a
Philip Leverhulme Prize for JMR from the Leverhulme Trust. JMR would also
like to thank the President and Fellows of Wolfson College, Cambridge, and
Prof John Young of the Engineering Department, Cambridge University, for
their support and hospitality during a sabbatical year when this work
was completed.

\bibliography{macro-continuum}
\bibliographystyle{elsart-num}

\appendix

\section{Further comments on the definitions of number-density and mass-density}
\label{numb-dens-app}
\subsection{Preliminaries}
Let us consider a fixed inertial reference frame, $\mathfrak{R}_G$,
with reference to the coordinate elements $(X_1,X_2,X_3)$. In this
fixed reference we investigate the motion of a cubic volume element
of fluid. We suppose that our cubic element of fluid is always
attached to a moving reference frame, $\mathfrak{R}_F$, with
coordinate elements $(F_1,F_2,F_3)$. Moreover, our cubic element of
fluid is determined at any time by the three base vectors of the
reference frame $\mathfrak{R}_F$.

For simplicity we assume that initially both reference frames
coincide, i.e., initially the three base vectors of both frames
$\mathfrak{R}_G$ and $\mathfrak{R}_F$ are the same.
Regarding the fixed frame $\mathfrak{R}_G$, the second frame
representing our element of fluid can have the following types of
motions: translation, rotation (according to three Eulerian angles),
and expansion or compression.

An element of volume in the fixed reference frame is denoted by
$d_X=d_{X_1}d_{X_2}d_{X_3}$ while an element of volume in the frame
representing the cubic element of fluid is
$d_F=d_{F_1}d_{F_2}d_{F_3}$. The element of volume $d_X$ and $d_F$
may be formally linked by  relation
\begin{equation}\label{jacobien}
   d_X =  J d_F ,
\end{equation}
where $ J$ is the absolute value of the Jacobian determinant of the
transformation of $\mathfrak{R}_G$ into $\mathfrak{R}_F$. If
$\mathfrak{R}_F$ is undergoing only translations or rotations in the
fixed reference $\mathfrak{R}_G$, then $J = 1$ as rotation and
translation conserve volumes. Otherwise, if compression or
expansion occurs then we have $ J \neq 1 $.

\subsection{The number density, $n$, and the mass-density, $\bar{\rho}$}
In our new kinetic approach introduced in this paper, $f(t, X, \xi,v
)$ is a probability number density in the phase space $(X, \xi, v)$.
This means the number of particles within an element of volume of
this phase space, $d_X d_\xi d_v$, is $f(t, X, \xi,v )d_X d_\xi
d_v$. The number of particles around a position, $X$, may be denoted
$n(t,X)d_X$ :
\begin{equation}\label{part-arr-x}
n(t,X)d_X  =   \int_{-\infty}^{+\infty}  \int_0^{+\infty} f(t, X,
\xi, v )  d_\xi d_v .
\end{equation}
The total number of gaseous particles in the whole space referenced
by the fixed frame $\mathfrak{R}_G$ is
\begin{equation}
N =  \int  n(t,X) d_X \ .
\end{equation}
Accordingly, $n(t,X)$ appears as a ``number-density'' referring to
the fixed reference frame $(X_1,X_2,X_3)$. It is a number of
particles divided by the fixed element of volume $d_X$.

Because of expansion or compression of the reference frame
$(F_1,F_2,F_3)$ during motion, we cannot presume a correspondence
between the element of volume $d_X$ and the element of volume $d_F$
which gives the actual volume of the fluid. Therefore, the correct
number-density referring to the real volume occupied by the fluid is
not directly given by $n$ but may be different by the dilatation or
compression coefficient $J$.

As the mean value of the microscopic volume of fluid around the
particles, $\bar{v}$, is defined through
\begin{equation}
n(t,X) \bar{v} (t,X) = \int_{-\infty}^{+\infty}  \int_0^{+\infty} v
f(t, X, \xi, v )  d_v d_\xi \ ,
\end{equation}
it is found, using equation (\ref{part-arr-x}), that the volume
occupied by the fluid at reference position $X$ is given by
$(nd_X)\bar{v}$, i.e. the total number of microscopic volumes of
fluid (or number of particles) multiplied by the mean value of these
microscopic volumes, $\bar{v}$. This volume of fluid contains $nd_X$
gaseous molecules. Therefore, when we refer to the actual volume of
the fluid, the ``number-density'' is
\begin{equation}
\label{volumedensitydef} \frac{n d_X}{n\bar{v}d_X} =
\frac{1}{\bar{v}} \ .
\end{equation}
The fluid density (or mass-density) in its physical meaning at
position $X$ is then
\begin{equation}
\label{massdensitydefAppend} \bar{\rho}  =  \frac{M n  d_X
}{n\bar{v}d_X} = \frac{M}{\bar{v}} \ .
\end{equation}

\subsection{The classical kinetic theory view of density}
In classical kinetic theory, molecules are assumed to always occupy,
and be uniformly dispersed in, the element of space $d_X$ without a
clear distinction between the actual volume occupied by the gaseous
molecules and the physical space connected to the fixed inertial
reference frame.  Chapman \& Cowling \cite{chapman} state that:
``\ldots the mass contained by $d_X$ will be proportional only to
its volume, and will not depend on its shape\ldots Similarly, the
number of molecules in $d_X$ \ldots is proportional to $d_X$. It
will be denoted by $n d_X$; $n$ is called the number-density of
molecules.''

In others words, $n d_X$ is the number of particles in the volume
element $d_X$ which itself is regarded as the volume of the fluid.
Therefore, in this description expansion or compression of the
volume element $d_F$ in the fixed reference $(X_1,X_2,X_3)$ is
disregarded.

According to this classical view, the number-density referring to
the element of volume $d_X$ is written
\begin{equation}
\label{volumedensitydefcla}
 \frac{n d_X }{d_X}  = n \ ,
\end{equation}
while the volume of fluid around each particle is given by
\begin{equation}
\label{specvolclass}  \frac{d_X}{nd_X}   = \frac{1} {n} \  .
\end{equation}
We see that the classical description imposes directly a unit volume
of fluid given by the inverse of the number density $n$. From our
equation (\ref{volumedensitydef}), the unit volume of fluid is
$\bar{v}$, which refers to the actual volume of the fluid, $d_F$.
The classical unit volume, $1/n$, refers to the volume element $d_X$ of the
fixed inertial reference frame. These quantities may therefore
differ by a compression/dilatation coefficient $J$.

If the particles are uniformly distributed, i.e.\ no mass density
variation in the fluid, then $n d_X$ is the number of particles
always occupying uniformly the volume $d_X$. In this case, the mean
microscopic volume, $\bar{v}$, around each particle is given by the
volume $d_X$ divided by the number of particles $nd_X$,
\begin{equation}
\label{unifo-v=n}
 \bar{v} =\frac{d_X} {nd_X} = \frac{1} {n} .
\end{equation}
Consequently, only in this particular situation does the definition
of the mass-density, equation (\ref{massdensitydefAppend}), become
\begin{equation}
\label{massdensitydef-uni} \bar{\rho} = \frac{M}{\bar{v}} = M n \ .
\end{equation}
Generally, however, $Mn$ is  different from $\bar{\rho}$. The
density quantity $Mn$ may be regarded as the density of a similar
fluid under similar conditions although with compression and
expansion motions removed; the actual density of the fluid is given
by $\bar{\rho}$. The product $n\bar{v}$ behaves like a
compression/dilatation coefficient, which depends on time and
position.

Finally, the distinction between the actual element of volume of
fluid and the simple element of volume taken in the fixed inertial
reference frame raises a question of consistency in the definition of
the pressure tensor from kinetic theory. The pressure tensor
definition invokes a volume element of fluid for which the classical
conceptual frame does not make a distinction between the fixed
element, $d_X$, and the element of volume of fluid, $d_F$. The real
element of fluid should therefore be decoupled from the fixed frame
of the observer.

\section{The Euler form of the continuity equation \label{euler-contra}}

Here we consider incompressible flows, by which we mean only $\nabla
\cdot U_m =0$. We suppose, however, that variations of density, and
then variations of the volume, of a fluid element can exist; for
example, through variations of temperature in time and space.

Let us follow a fixed amount of mass, $M_f$, of fluid, occupying at
time $t_0$ a volume $V_0$, and occupying at time $t$ the volume $V$.
In classical fluid mechanics, if we denote $\rho\equiv \rho(t,X)$
the density of the fluid, then we should have
\begin{equation}
\label{euler0} M_f = \int_{V_0} \rho_0 d_{V_0}   =   \int_V \rho d_V
\ .
\end{equation}
A change of variables can be applied so that
\begin{equation}
\label{euler1}
 \int_{V_0} \rho_0 d_{V_0}   =   \int_{V_0} \rho J d_{V_0} \ ,
\end{equation}
where, as in equation (\ref{jacobien}), $J$ is due to the
application transforming $V_0$ into $V$. Equation (\ref{euler1}) may
be applied at $t = t_0$, in which case $J(t=t_0)=J_0=1$. So equation
(\ref{euler1}) can be written
\begin{equation}
\label{euler2}
 \int_{V_0} [J_0\rho_0 - J\rho]d_{V_0}   = 0 \ .
\end{equation}
As $M_f$ is an arbitrary amount of mass, and $V_0$  an arbitrary
volume, equation (\ref{euler2}) becomes
\begin{equation}
\label{euler3}
  [J_0\rho_0 - J\rho]   = 0 \ ,
\end{equation}
so we can write the following:
\begin{equation}
\label{euler4}
  \frac{D ( J\rho)}{ D t}  = 0 \ ,
\end{equation}
for any time $t$. In equation (\ref{euler4}) the derivative is a
material (convective) derivative because we are explicitly moving
with the mass, as we impose $M_f$ to be constant. This equation, due
to Euler, is known as the material form of the continuity equation
\cite{malvern}. We note that in our case $J$ depends on time $t$: at
$t=t_0$, $J=1$, but nothing is stated about the derivative of $J$
(i.e., nothing imposes $D J / Dt = 0 $ for $t=t_0)$.

The local form of the conventional continuity equation (with $\nabla
\cdot U_m =0$) is
\begin{equation}
\label{local}
  \frac{D \rho}{ D t}  = 0 \ .
\end{equation}
Therefore, a contradiction appears between the material form of the
continuity equation (\ref{euler4}) and the common fluid mechanics
expression (\ref{local}) because $J$ depends on time --- but these
two expressions should be the same as they are expressing the same
physical law.

Using our description of density from Appendix \ref{numb-dens-app},
the density of the fluid which should be used in equations
(\ref{euler0})--(\ref{euler4}) is $\rho= \bar{\rho} = M/\bar{v}$,
where the expansion/compression coefficient $J$ is $n\bar{v}$. So,
replacing these elements in equation (\ref{euler4}), we find
\begin{equation}
\label{localmoi}
  \frac{D (M n)}{ D t}  = 0 \ ,
\end{equation}
which is the correct local form of the mass continuity equation: the
contradiction existing between the Euler equation (\ref{euler4}) and
the classical local form of the continuity equation disappears in
our description in which the density is $\bar{\rho}$, equation
(\ref{massdensitydefAppend}), and the local form of the continuity
equation is equation (\ref{localmoi}). The Euler equation
(\ref{euler4}) and equation (\ref{localmoi}) are therefore entirely
equivalent in our description.

The density of the fluid, $\bar{\rho}$, which varies according to
changes in the properties of the fluid, satisfies the Euler form of
the continuity equation; the quantity $Mn$ behaves like a reference
density, retracing the conservation of mass from a reference frame
in which any change in the properties of the fluid is observed. Our
reference amount of mass, $M_f$, is simply always constant.

\end{document}